\documentclass[12pt]{article}
\usepackage{amssymb,amsmath,epsfig}

\begin{document}

\title{\bf Phantom Accretion by Five Dimensional Charged Black Hole}

\author{M. Sharif \thanks{msharif@math.pu.edu.pk} and G. Abbas
\thanks{abbasg91@yahoo.com}\\
Department of Mathematics, University of the Punjab,\\
Quaid-e-Azam Campus, Lahore-54590, Pakistan.}

\date{}
\maketitle

\begin{abstract}
This paper deals with the dynamical behavior of phantom field near
five dimensional charged black hole. We formulate equations of
motion for steady-state spherically symmetric flow of phantom
fluids. It is found that phantom energy accretes onto black holes
for $u<0$. Further, the location of critical point of accretion are
evaluated that leads to mass to charge ratio for $5D$ charged black
hole. This ratio implies that accretion cannot transform a black
hole into a naked singularity. We would like to mention here that
this work is an irreducible extension of $4D$ charged black hole.
\end{abstract}

\section{Introduction}

During the last decades, there has been a growing interest to study
the gravity in a theory which implies the existence of extra
dimension in nature, called brane-world theories. Such theories
suggest the solution of hierarchy problem (difference in scales of
gravitational and electro-weak interaction) \cite{1}. The
brane-world theories are based on the fact that (3+1)-dimensional
brane is embedded in a (4+$n$)-dimensional spacetime with $n$
spacelike compact dimension \cite{2}. All the matter is located on
brane and fields propagate in the bulk \cite{3}. There is a
possibility that brane-world gravity theory would give an idea to
observe the effects of quantum gravity in laboratory at TeV
energies. Also, these theories urge that higher dimensional black
hole (BH) can be produced in large hadron colliders (LHC) and cosmic
ray experiments \cite{4}.

The development of higher dimensional theories has increased the
interest to study BH in higher dimension. The first static
spherically symmetric BH solution in brane-world was formulated by
Dadhich et al. \cite{5}, which has the same structure as
Reissner-Nordsrt$\ddot{o}$m (RN) $4D$ BH solution. Its physical
interpretation implies that there is a tidal charge due to the fifth
dimension. Konoplya and Zhidenko \cite{3} discussed the higher
dimensional charged BH instability. Guha et al. \cite{6}
examined the geodesic motion in the vicinity of $5D$ RN anti
de-Sitter BH. Ghosh et al. \cite{7} introduced the idea of
gravitational collapse in $5D$ for the dust case. This was extended
by Sharif and his collaborators for perfect fluid \cite{8} and
charged perfect fluid \cite{9} collapse in $5D$. Matzner and
Mezzacappa \cite{10} examined the closed universes in $5D$
Kaluza-Klien theory. These studies motivate us to explore the
problem of phantom accretion in $5D$ static spherically symmetric
charged BH solution which is characterized by mass and electric
charge.

It was first confirmed by the data of type $Ia$ Supernova and
large scale structure \cite{11}-\cite{14} that our universe is in
accelerating phase. Different models \cite{15}-\cite{18} were
proposed to understand the nature of DE in our universe. The
simplest form of DE is vacuum energy (cosmological constant) for
which the equation of state parameter (EoS) is $\omega=-1$. The
quintessence and phantom are the forms of DE for which $\omega>-1$
and $\omega<-1 $, respectively \cite{19}-\cite{21}. Phantom energy
violates the dominant energy condition.

The problem of matter accretion onto the compact objects in
Newtonian gravity was first formulated by  Bondi \cite{22}. In
general relativity, Michel \cite{23} was the pioneer who studied
accretion of gas onto the Schwarzschild BH. Sun \cite{24} discussed
the phantom accretion onto BH in the cyclic universe. Babichev et
al. \cite{25} have shown that BH mass diminishes due to phantom
accretion. Jamil et al. \cite{26} have explored the effects of
phantom accretion onto the charged BH in $4D$. They pointed out that
if mass of BH becomes smaller (due to accretion of phantom energy)
than its charge, then Cosmic Censorship Hypothesis is violated.

In this paper, we extend this work for phantom accretion by $5D$
charged BH. We find that accretion cannot transform a BH into a
naked singularity or extremal BH, in contrast to $4D$ case. The plan
of the paper is as follows: In the next section, accretion onto $5D$
charged BH is presented. We discuss the critical accretion in
section \textbf{3} and conclude our discussion in the last section.
The gravitational units (i.e., the gravitational constant $G=1$ and
speed of light in vacuum $c=1$) are used. All the Latin and Greek
indices vary from 0 to 4, otherwise it will be mentioned.

\section{Accretion onto $5D$ Charged Black Hole}

We consider a charged static spherically symmetric $n+2$ dimensional
BH solution \cite{3}
\begin{equation}\label{1}
ds^2=Z(r)dt^2-\frac{1}{Z(r)}dr^2-r^2d\Omega_n,
\end{equation}
where $d\Omega_n$ is the unit $n$ sphere and
$Z(r)=1-\frac{2M}{r^{n-1}}+\frac{Q^2}{r^{2n-2}}$. For $n=2$, this
reduces to $4D$ RN metric, while for $n=3$, we get a $5D$ charged BH
solution given by
\begin{equation}\label{2}
ds^2=Z(r)dt^2-\frac{1}{Z(r)}dr^2-r^2(d\theta^2+\sin^2{\theta}d\phi^2
+\sin^2{\theta}\sin^2{\phi}d\psi^2),
\end{equation}
where $Z(r)=1-\frac{2M}{r^2}+\frac{Q^2}{r^4}$. Here $M$ and $Q$ are
the mass and charge of the BH.

The black hole horizons can be found by solving
$Z(r)=1-\frac{2M}{r^2}+\frac{Q^2}{r^4}\equiv 0$, for $r$ whose
positive real roots will give horizons as follows
\begin{eqnarray}\label{2}
r_{outer}=\sqrt{ M+{\sqrt{M^2-Q^2}}},\quad r_{inner}=\sqrt{
M-{\sqrt{M^2-Q^2}}}.
\end{eqnarray}
For $M^2>Q^2,~r_{outer}>r_{inner}$, for
$M^2=Q^2,~r_{outer}=r_{inner}\equiv m$ (an extreme charged BH) and
for $M^2<Q^2$, both horizons disappear and singularity becomes naked
at $r=0$. For $Q=0,~r_{outer}=2m$ (Schwarzschild horizon in $4D$)
and $r_{inner}=0$. This implies that like $4D$ case, the existence
of charge is necessary for the existence of inner horizon (Cauchy
horizon). The regularity of the $5D$ charged BH can be seen in the
regions $r_{outer}<r<\infty,~r_{inner}<r<r_{outer}$ and
$0<r<r_{inner}$.

The energy-momentum tensor for phantom energy is
\begin{equation}\label{3}
{T_{{\mu}{\nu}}={({\rho}+p)}u_{\mu}u_{\nu}-pg_{\mu\nu}},
\end{equation}
where $\rho$ is the energy density, $p$ is the pressure and
$u^\mu=(u^t,u^r,0,0,0)$ is the five-vector velocity. It is mentioned
here that $u^\mu$ satisfies the normalization condition, i.e.,
$u^\mu u_\mu =-1$. The conservation of energy-momentum tensor yields
\begin{equation}\label{4}
r^2u{M}^{-2}(\rho+p)\left(Z(r)+u^2\right)^{\frac{1}{2}}=C_0,
\end{equation}
where $C_0$ is an integration constant and $u^{r}=u<0$ for inward
flow.

The energy flux equation can be derived by projecting the
energy-momentum conservation law on the five-velocity, i.e.,
${u_\mu T^{\mu\nu}}_{;\nu}$=0 for which Eq.(\ref{3}) leads to
\begin{equation}\label{5}
r^2u{M}^{-2}\exp
\left[\int^{\rho_h}_{\rho_\infty}\frac{d\rho'}{\rho'+p(\rho')}\right]=C_1,
\end{equation}
where $C_1>0$ is another integration constant which is related to
the energy flux. Also, $\rho_h$ and ${\rho_\infty}$ are densities of
the phantom energy at horizon and at infinity. From Eqs.(\ref{4})
and (\ref{5}), it follows that
\begin{equation}\label{6}
(\rho+p)\left(Z(r)+u^2\right)^{\frac{1}{2}}
\exp\left[-\int^\rho_{\rho_\infty}\frac{d\rho'}{\rho'+p(\rho')}\right]=C_2,
\end{equation}
where $C_2=-\frac{C_0}{C1}=\rho_\infty +p(\rho_\infty)$.

The rate of change of BH mass due to fluid accretion onto it is
\cite {27}
\begin{equation}\label{7}
\dot{M}=-4 \pi r^2 {T^r}_t.
\end{equation}
Using Eqs.(\ref{4})-(\ref{6}) in the above equation, it follows
that
\begin{equation}\label{8}
\dot{M}=4\pi M^2C_1({\rho}_\infty +{p}_\infty).
\end{equation}
We note that mass of BH decreases if $({\rho}_\infty
+{p}_\infty)<0$. Thus the accretion of phantom energy onto a BH
causes to decrease the mass of BH. As the phantom accretion only
diminishes mass and does not affect the charge of BH, so we can
speculate that when $M^2<Q^2$ is reached, then singularity becomes
naked at $r=0$ and the phantom accretion by $5D$ charged BH may lead
to the violation of Cosmic Censorship Hypothesis. However, critical
accretion process mentioned below implies that Cosmic Censorship
Hypothesis remains valid in this case. It is mentioned here that one can
solve Eq.(\ref{8}) for $M$ by using EoS $p=k\rho$. Since all $p$ and
$\rho$, violating dominant energy condition, must satisfy this
equation, hence it holds in general. i.e.,
\begin{equation}\label{01}
\dot{M}=4\pi M^2C_1({\rho}+{p}).
\end{equation}

\section{Critical Accretion}

Here, we locate such points at which flow speed is equal to the
speed of sound during accretion. The fluid falls onto the BH with
monotonically increasing velocity along the particle trajectories.
We follow the procedure introduced by Michel \cite {23}. The
conservation of mass flux, ${J^\mu}_{;~\mu}=0$, gives
${\setcounter{equation}{0}}$
\begin{equation}\label{9}
\rho u r^2=k,
\end{equation}
where $k$ is an integration constant. From Eqs.(\ref{4}) and
(\ref{8}), we get
\begin{equation}\label{10}
\left(\frac{\rho +p}{\rho}\right)^2 \left( Z(r)+u^2\right)=k_1,
\end{equation}
where $k_1=(\frac{C_0}{k})^2$ is a positive constant.
Differentiating Eqs.(\ref{9}) and (\ref{10}) and eliminating
$d\rho$, we get
\begin{equation}\label{11}
\frac{dr}{r}\left[2V^2-\frac{\frac{M}{r^2}-\frac{Q^2}{r^4}}{Z(r)+u^2}\right]+\frac{du}{u}
\left[V^2-\frac{u^2}{Z(r)+u^2}\right]=0,
\end{equation}
where $V^2=\frac{dln(\rho+p)}{dln\rho}$. This shows that critical
points are found by taking both the factors inside the square
brackets equal to zero. Thus we obtain
\begin{equation}\label{12}
{u_\ast}^2=\frac{M{r_\ast}^2-{Q}^2}{r_\ast^4},\quad
{V_\ast}^2=\frac{M{r_\ast}^2-{Q}^2}{{r_\ast}^4-M{r_\ast}^2}.
\end{equation}
We see that physically acceptable solutions of the above equations
are obtained if ${u_\ast}^2>0$ and ${V_\ast}^2>0$ implying that
\begin{eqnarray}\label{14}
M{r_\ast}^2-{Q}^2>0, \quad {r_\ast}^4-M{r_\ast}^2>0.
\end{eqnarray}
The subscript $\ast$ is used to represent a quantity at a point
where speed of flow is equal to the speed of sound, such a point is
called a critical point. It is mentioned here that in case of $4D$
charged BH hole the equations corresponding to Eq.(\ref{14}) are
linear and quadratic in $r$.

The solution of the 2nd equation of Eq.(\ref{14}) is
\begin{equation}\label{16}
r_{\ast_+}>\sqrt{M}.
\end{equation}
For the solution about critical point, we insert the value of
$r_{\ast+}$ in the first equation of Eq.(\ref{14}) and obtain
\begin{equation}\label{17}
1<\frac{M^2}{Q^2}.
\end{equation}
Thus accretion through $r_{\ast_+}$ is possible if the above mass to
charge ratio is satisfied. Since horizons always exist for this mass
to charge ratio, so in contrast to $4D$ charged BH there are
no possibilities of extremal BH and naked singularity
formation during the accretion process.

\section{Outlook}

In this paper, we have analyzed the phantom accretion by $5D$
charged BH. We have formulated equations of motion for steady state
spherically symmetric phantom flow near $5D$ charged BH. It has been
assumed that infalling fluid does not disturb the generic properties
of the BH. Following the procedure introduced by Michel \cite{23},
we discuss the accretion and critical accretion by BH. Like the
cases of $4D$ Schwarzschild and RN, phantom accretion decreases the
mass of BH.

Two (event and Cauchy) horizons for $5D$ charged BHs can
exist only if $M^2\geqslant Q^2$ otherwise there will be a naked
singularity. We have found that the existence of Cauchy horizon
requires $Q\neq0$. If we take $Q=0$ then there exists a unique
horizon which is at $r=2m$ ($4D$ Schwarzschild radius). The critical
accretion analysis implies that corresponding to two horizons there
exists a positive value of $r_\ast$ (i.e., $r_{\ast_+}>\sqrt{m}$).
This can play the role of physically possible critical point if the
mass and charge of $5D$ BH satisfies $1<\frac{M^2}{Q^2}$. In
contrast to $4D$ charged BH case, this ratio is free of upper bound.
Further, this ratio implies that $M^2>Q^2$, which is essential
inequality for the existence of horizons. It is concluded that
although phantom accretion decreases the mass of BH, but it cannot
be converted to $M^2\leq Q^2$. Hence throughout the accretion
process, a $5D$ charged BH cannot be transformed to an extremal
charged BH or a naked singularity and Cosmic Censorship remains
valid in this case.

\newpage
\vspace{0.5cm}

{\bf Acknowledgment}

\vspace{0.25cm}

We would like to thank the Higher Education Commission, Islamabad,
Pakistan for its financial support through the {\it Indigenous Ph.D.
5000 Fellowship Program Batch-IV}.

\end{document}